\documentclass[iop]{emulateapj}

\usepackage{graphicx}
\usepackage{amsmath}

\slugcomment{Accepted on March 27, 2016}

\shorttitle{How Lyman Alpha Emission Depends on Galaxy Stellar Mass}
\shortauthors{Oyarz\'un et al.}

\begin{document}


\title{How Lyman Alpha Emission Depends on Galaxy Stellar Mass}


\author{Grecco A. Oyarz\'un\altaffilmark{1},  Guillermo A. Blanc\altaffilmark{1,2,3}, Valentino Gonz\'alez\altaffilmark{1,2}, Mario Mateo\altaffilmark{4}, John I. Bailey III\altaffilmark{4}, 
Steven L. Finkelstein\altaffilmark{5}, Paulina Lira\altaffilmark{1}, 
Jeffrey D. Crane\altaffilmark{6}, 
and Edward W. Olszewski\altaffilmark{7}}

\affil{$^{1}$Departamento de Astronom\'ia, Universidad de Chile, Casilla 36-D, Santiago, Chile}
\affil{$^{2}$Centro de Astrof\'isica y Tecnolog\'ias Afines (CATA), Camino del Observatorio 1515, Las Condes, Santiago, Chile.}
\affil{$^{3}$Visiting Astronomer, Observatories of the Carnegie Institution for Science, 813 Santa Barbara St, Pasadena, CA, 91101, USA}
\affil{$^{4}$Department of Astronomy, University of Michigan, Ann Arbor, MI 48109, USA}
\affil{$^{5}$Department of Astronomy, The University of Texas at Austin, Austin, TX 78712, USA}
\affil{$^{6}$The Observatories of the Carnegie Institution for Science, 813 Santa Barbara Street, Pasadena, CA 91101, USA}
\affil{$^{7}$Steward Observatory, University of Arizona, Tucson, AZ, USA}



\begin{abstract}
In this work, we show how the stellar mass ($M_*$) of galaxies affects the $3<z<4.6$ Ly$\alpha$ equivalent width (EW) distribution. To this end, we design a sample of 629 galaxies in the $M_*$ range $7.6<\log{M_*/M_{\odot}}<10.6$ from the 3D-HST/CANDELS survey. We perform spectroscopic observations of this sample using the Michigan/Magellan Fiber System, allowing us to measure Ly$\alpha$ fluxes and use 3D-HST/CANDELS ancillary data. In order to study the Ly$\alpha$ EW distribution dependence on $M_*$, we split the whole sample in three stellar mass bins. We find that, in all bins, the distribution is best represented by an exponential profile of the form $dN(M_{*})/dEW=W_{0}(M_{*})^{-1}A(M_{*})e^{-EW/{W_{0}(M_{*})}}$. Through a Bayesian analysis, we confirm that lower $M_*$ galaxies have higher Ly$\alpha$ EWs. We also find that the fraction $A$ of galaxies featuring emission and the e-folding scale $W_{0}$ of the distribution anti-correlate with $M_*$, recovering expressions of the forms $A(M_*) = -0.26(.13)\log{M_*/M_{\odot}}+ 3.01(1.2)$ and $W_{0} (M_*) = -15.6(3.5) \log{M_*/M_{\odot}} + 166(34)$. These results are crucial for proper interpretation of Ly$\alpha$ emission trends reported in the literature that may be affected by strong $M_*$ selection biases.
\end{abstract}



\keywords{galaxies: evolution - galaxies: high-redshift - galaxies: statistics}


\section{Introduction}

A lot of progress has been made toward understanding the physics and statistics of Ly$\alpha$ emission at high redshift (e.g.\ Shapley et al.\ 2003, Ouchi et al.\ 2008, Stark et al.\ 2010, Blanc et al.\ 2011). Still, current estimations of the magnitude and frequency of this process are limited by biases that emerge from the sample selection techniques employed. For example, spectroscopic studies of UV continuum detected galaxies show that only about 50\% of Lyman break galaxies (LBGs) at $z=3$ feature Ly$\alpha$ in emission, while the other half shows absorption (Shapley et al.\ 2003, Stark et al.\ 2010). These studies also find an anti-correlation between the UV luminosities of galaxies and their Ly$\alpha$ equivalent widths (EWs), as well as a significant increase in the fraction of galaxies showing large Ly$\alpha$ EWs (e.g.\ $>75$\AA) when going from $z\sim3$ to
$z\sim6$ (Stark et al.\ 2010). On the other hand, narrowband imaging selected samples of Ly$\alpha$ emitters (LAEs) include, by construction, only objects showing Ly$\alpha$ above a certain EW detection threshold. Nevertheless, 
even in this regime, significant differences are seen with respect to the statistics derived from high Ly$\alpha$ EW LBG samples. For instance, the EW distribution of LAEs does not seem to evolve significantly over the $3<z<6$ range (Ouchi et al.\ 2008, Zheng et al.\ 2014), while it does seem to shift toward lower EWs at lower redshifts ($z\sim2$, Ciardullo et al.\ 2012). Furthermore, unlike LBGs, LAEs show very little correlation between their UV luminosities and their Ly$\alpha$ EWs (Ouchi et al.\ 2008).

The reason for some of these discrepancies lies in the fact that different high-redshift galaxy selection techniques sample different regions of the stellar mass ($M_{*}$), star formation rate, and metallicity parameter space. These parameters can affect the production and escape of Ly$\alpha$ photons through correlations with stellar population ages, neutral hydrogen mass, and dust abundance. High-mass stars present in stellar populations younger than 10 Myr are responsible for Ly$\alpha$ emission, with the effect decreasing as these populations grow older (Charlot \& Fall 1993, Schaerer 2003). Ly$\alpha$ radiative transfer is also severely 
affected by the neutral gas structure and kinematics of the ISM and circumgalactic medium (Verhamme et al.\ 2006). Likewise, the Ly$\alpha$ escape fraction is known to strongly anti-correlate with dust extinction, at least for high-EW objects (Blanc et al.\ 2011, Hagen et al.\ 2014). Considering that more massive galaxies tend to have older stellar populations, higher gas mass, and more dust in their ISM, Ly$\alpha$ emission is severely affected by $M_*$. If we take into account that emission line surveys sample a lower range in $M_*$ than LBGs samples due to the first not requiring a continuum detection, Ly$\alpha$ statistics are highly dependent on survey design and $M_*$ completeness.

In order to assess the effects of $M_*$ on high-redshift Ly$\alpha$ emission, we present a spectroscopic survey of an $M_*$ selected sample of $3<z<4.6$ galaxies. We conduct this survey with M2FS (Mateo et al.\ 2012) at the Magellan-II Clay telescope. Using a Bayesian approach, we quantify the $3<z<4.6$ Ly$\alpha$ EW distribution dependence on $M_*$. This Letter is structured as follows. In Section 2, we describe our sample and data set. In Sections 3 and 4, we explain our methodology and results. Implications are presented in Section 5. We adopt a $\Lambda$CDM cosmology with $H_{0}$ = 70 km s$^{-1}$ Mpc$^{-1}$, $\Omega_{m}$ = 0.3, and $\Omega_{\lambda}$ = 0.7.

\section{Data}
\subsection{Sample Selection}
Our sample is composed of 629 galaxies in the COSMOS, GOODS-S and UDS fields. Every object is observed under the 3D-HST/CANDELS program (Grogin et al.\ 2011; Koekemoer et al.\ 2011), providing HST/Spitzer photometry from 3800 \AA \ to 7.9 $\mu m$ (44 bands for COSMOS, 40 for GOODS-S, and 18 for UDS). We construct our sample using 3D-HST outputs (Skelton et al.\ 2014). According to these, our 629 photometric redshifts satisfy $3.25<z_{3D-HST}<4.25$ and have a 95\% probability of $2.9<z<4.25$. Every galaxy also complies with a photometric redshift reliability parameter $Q_{z}\leqslant3$ selection to remove catastrophic outliers (Brammer et al.\ 2008). In terms of $M_*$, our galaxies are homogeneously distributed in the range $8<\log({M_*/M_{\odot}})_{3D-HST}<10.4$. These values are obtained assuming exponentially declining star formation histories (SFHs) with a minimum e-folding time of $log_{10}(\tau/year) = 7$ (Skelton et al.\ 2014).

\subsection{Observations}
Spectroscopy of the complete sample was conducted at the Magellan Clay 6.5 m telescope during 2014 December and 2015 February. To this end, we used M2FS, a multi-object fiber-fed spectrograph. This instrument's 1.$''$2 fibers allow for 256 targets, of which we used 40 for sky apertures. The final data set consists of six exposure hours on each of the three fields with an average seeing of 0.$''$6.

\begin{figure}
	\centering
	\includegraphics[width=3.4in]{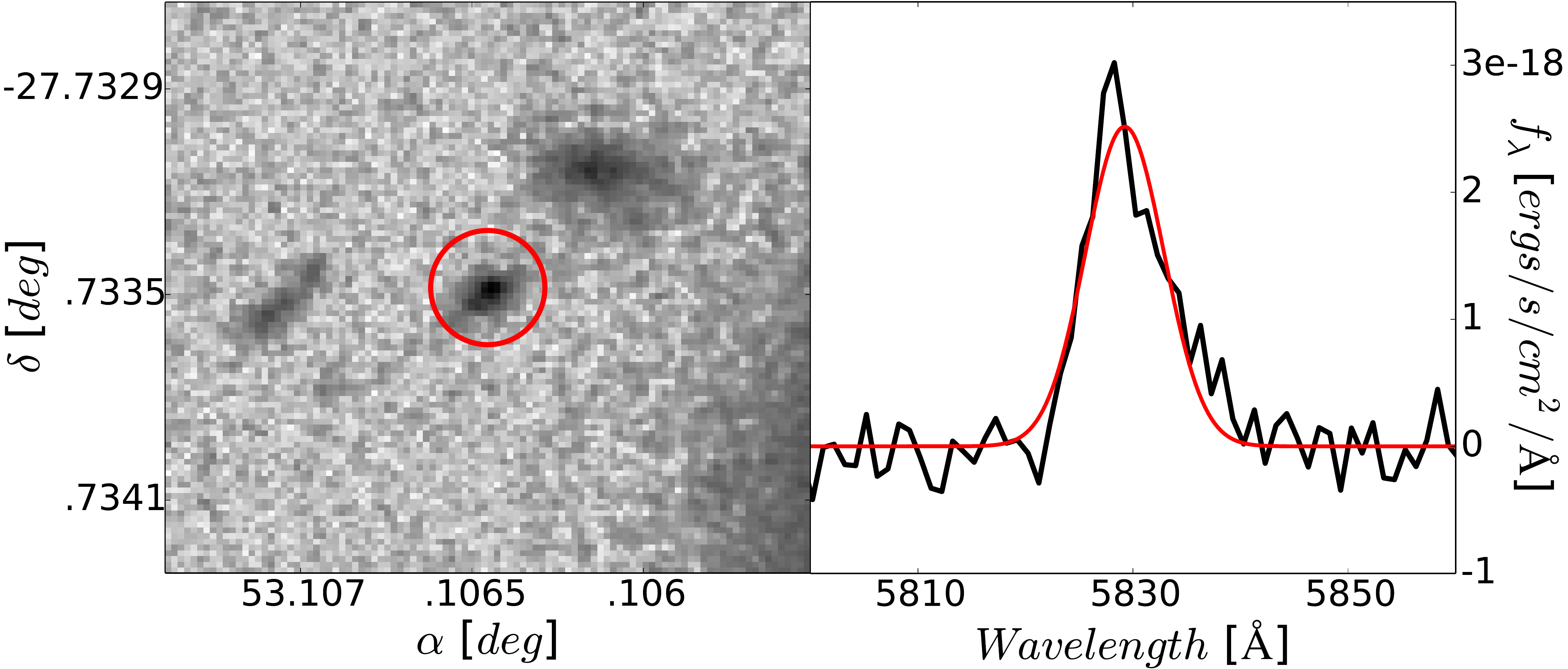}
	\caption{Left: CANDELS F160W image for target GS38014 ($m_{F160W}=24.5$), along with M2FS 1.2$''$ fiber in red. Right: reduced spectrum of GS38014 with our best gaussian fit in red. We measure a Ly$\alpha$ flux of $2.3\times10^{-17}$ erg s$^{-1}$ cm$^{-2}$, $z_{Ly\alpha}\sim 3.795$, and $EW\sim84$ \AA \ for this line.}	
\end{figure}

\begin{figure}[h!]
	\centering
	\includegraphics[width=3.25in]{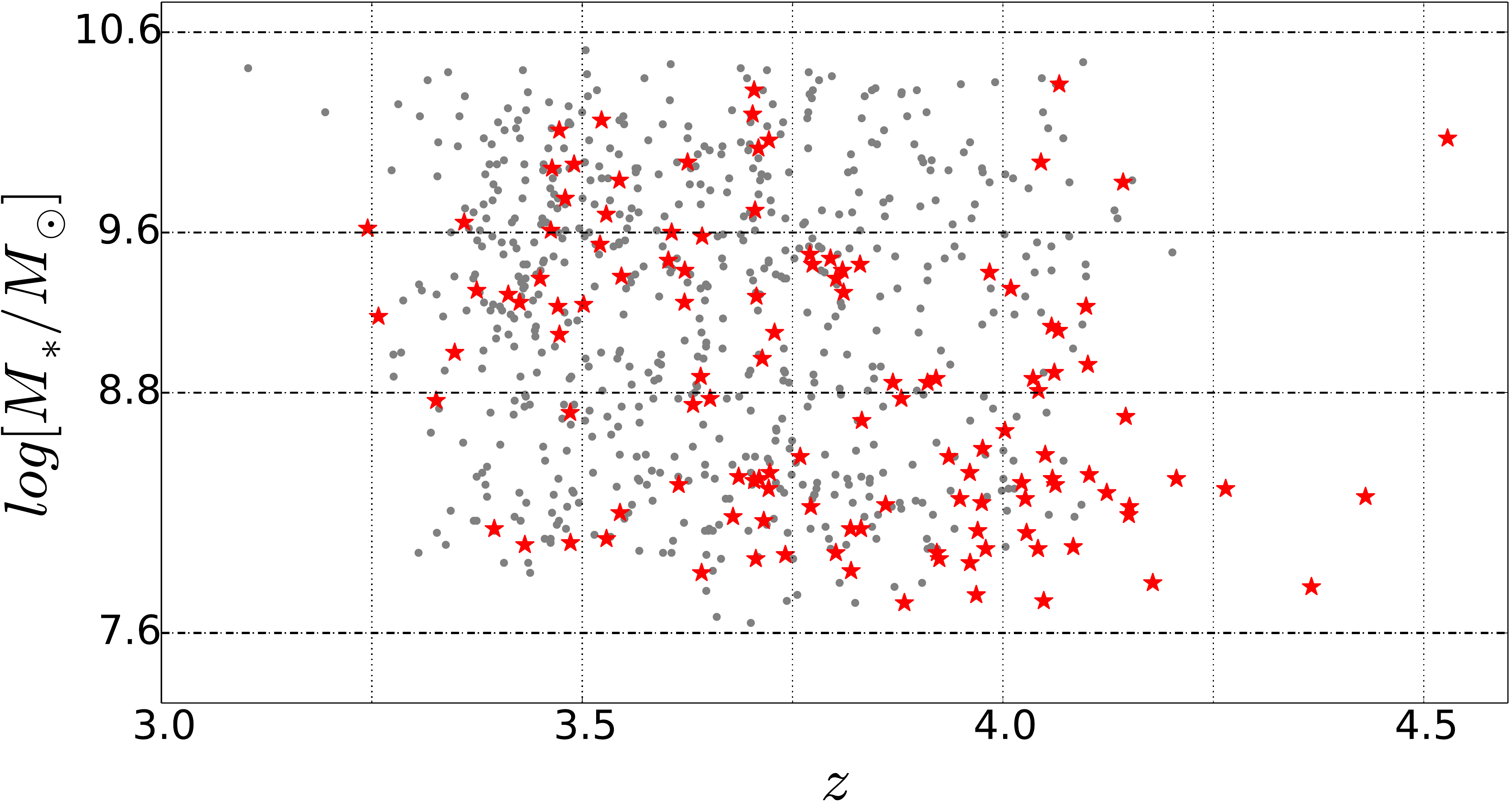} 
    \caption{Complete sample distribution in $z_{EAZY}$ and $M_*$, according to our EAZY and FAST outputs (dots). Overlaid are our three $M_*$ bins with median masses $\log{M_*/M_{\odot}}=8.3$, $9.3$, $10$. The red stars show our 120 spectroscopic Ly$\alpha$ redshifts ($SN\geqslant 5.5$). Note that detections are plotted twice (as  $z_{EAZY}$ and $z_{Ly\alpha}$).}
	\label{FigVibStab}
	\label{FigVibStab}
\end{figure}

Data reduction features standard bias subtraction, dark correction, wavelength calibration, flat-fielding, sky subtraction, and flux calibration. The resulting spectra FWHM line resolution is of $\sim 2$\AA. We reach a 1$\sigma$ continuum flux density limit of $\sim 4\times10^{-17}$ erg s$^{-1}$ cm$^{-2}$ \AA$^{-1}$ per pixel in our 6 hr of exposure. This translates into a 5$\sigma$ emission line flux sensitivity of $\sim 4\times10^{-18}$ erg
s$^{-1}$ cm$^{-2}$ in our final spectra. A sample galaxy with its reduced spectrum is shown in Figure 1. 

Flux calibration is performed using five $M_V=19-22$ calibration stars on each exposure, with an associated rms uncertainty of $\sim 15\%$. We are correcting for a $\sim32$\% fiber flux loss, which corresponds to a point-source Ly$\alpha$ surface brightness distribution.

\section{Methodology}

\subsection{Line detection}
We detect and characterize lines across the spectra using an automated maximum likelihood fitting routine. We assume intrinsic gaussian profiles of the form:
\begin{gather}
f_{rest}(\lambda)=\frac{f_{Ly\alpha}}{\sqrt{2\pi}\sigma_{\lambda}}e^{-(\lambda-\lambda_{0})^{2}/2\sigma_{\lambda}^{2}}
\end{gather}
where $f_{Ly\alpha}$, $\lambda_{0}$, and $\sigma_{\lambda}$ compose the parameter space explored by the maximum likelihood. Considering the resonant scattering and double-peaked nature of the Ly$\alpha$ line, gaussian profiles are just an approximation. Nevertheless, these are sufficient four our needs (Figure 1).

We run our line detection code on the 115 sky fibers to account for false positives. We detect four false lines above 4$\sigma$ and none above 5$\sigma$. Therefore, down to 5$\sigma$, we are confident of having less than 5 false detections in our $629$ targets. This translates into $\lesssim 5\%$ contamination using signal-to-noise (SN) $SN^{*}=5.5$ as our threshold, considering we have 120 detections with $SN \geqslant 5.5$ (Figure 2).

\subsection{Corrected  Parameters}
We run EAZY (Brammer et al.\ 2008) on CANDELS/IRAC photometry to obtain our own photometric redshifts ($z_{EAZY}$). The 629 objects satisfy $3<z_{EAZY}<4.25$ (Figure 2), with a median $\sigma_{EAZY}=0.1$. They also have a 95\% probability of $2.95<z<4.5$. From now on, we use our spectroscopic redshifts ($z_{Ly\alpha}$) for detections and $z_{EAZY}$ for non-detections. We find a median redshift offset of $\Delta z= z_{Ly\alpha}-z_{EAZY}=0.24$ for detections (Figure 2) and assess it in the conclusions. To have our own $M_*$ estimates, we run FAST (Kriek et al.\ 2009). This allows us to use constant SFHs and $z_{Ly\alpha}$ when available. Our outputs yield a mass coverage of $7.6<\log{M_*/M_{\odot}}<10.6$ (Figure 2), with a characteristic uncertainty of $\log{M_*/M_{\odot}}\sim 0.2$. We stress our galaxy sample does not feature any other selection cuts apart from possible photometric redshift biases and 3D-HST/CANDELS incompleteness, which is restricted to our low-mass bin ($\log{M_*/M_{\odot}}<8.5$). 
\begin{figure*}
	\centering
	\includegraphics[width=7in]{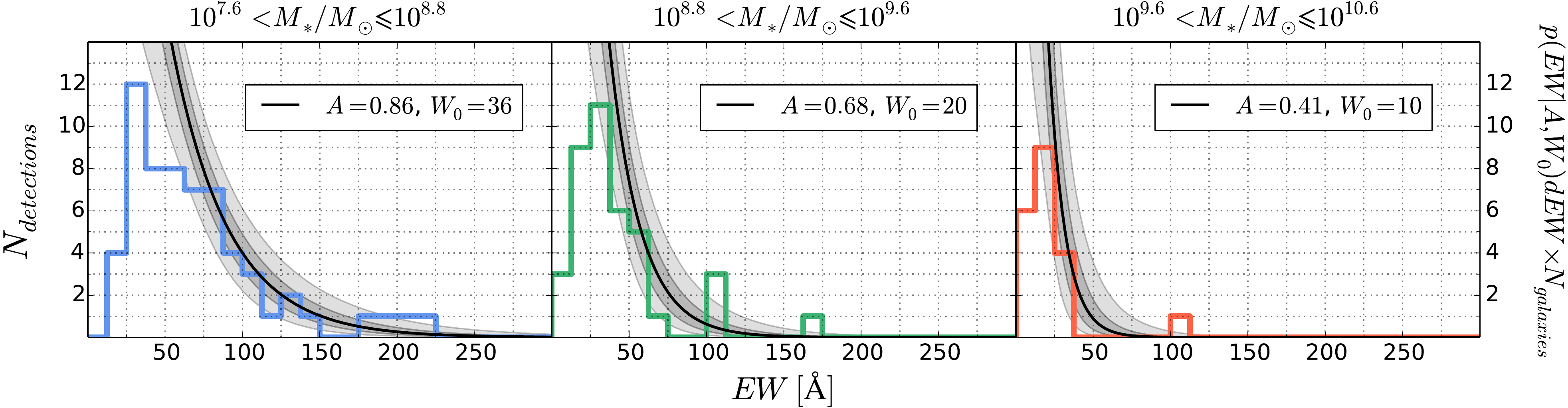}
	\caption{Rest frame Ly$\alpha$ EW distributions for the low-, medium-, and high-mass subsamples, respectively. Only detections with $SN\geqslant 5.5$ are shown. Overplotted are our best exponential distributions for every subsample (solid). Shaded are $1\sigma$ and $3\sigma$ constraints.}
	\label{FigVibStab}
\end{figure*}

\subsection{Bayesian Inference}
We calculate the Ly$\alpha$ EW in the rest frame as
\begin{equation}
EW=\frac{F}{f_{\lambda}}\frac{1}{(1+z_{Ly\alpha})}
\end{equation}
with $F$ the Ly$\alpha$ flux we measure in the spectra and $f_{\lambda}$ the observed flux at rest $1700$\AA \ from CANDELS photometry.

To reproduce Ly$\alpha$ EW distributions, the most widely used models are gaussian (e.g.\ Stark et al.\ 2010) and exponential (e.g.\ Zheng et al.\ 2014) profiles. We consider both functions and find the exponential to be the most appropriate to represent our distributions:
\begin{gather}
	p(EW|A, W_{0})=\frac{A}{W_{0}}e^{-EW/W_{0}}H(EW) +(1-A)\delta(EW)
		\label{dist}
\end{gather}
with $H(EW)$ the Heaviside and $\delta(EW)$ the Delta. Hence, $A$ is the fraction of galaxies featuring emission, $(1-A)$ the fraction of galaxies not showing emission, and  $W_{0}$ the e-folding scale of the distribution.

For convenience, we perform our Bayesian analysis using Ly$\alpha$ line flux $F$ instead of EW, as introduced in equation (2). According to Bayes's Theorem, the posterior distribution $p(A,W_{0}| \{{\mbox{$F$}}\} )$, i.e., the parameter space probability distribution given our data set $\{ F\}$, is
\begin{gather}
	p(A,W_{0}| \{ F\} )= \frac{p(\{ F\}|A,W_{0})p(A,W_{0})}{p(\{ F\} )}
	\label{bayes}
\end{gather}

As galaxies are independent, the likelihood is just the product of the individual likelihoods for every galaxy, i.e., $p(\{F\} |A,W_{0})=  \displaystyle \prod p(F_{i} |A,W_{0})$. For a galaxy with rest UV continuum flux $f_{\lambda,i}$ and uncertainty $\sigma_{\lambda, i}$, the single likelihood is given by
\begin{gather}
p(F_{i}|A, W_{0})  =
	\int_{0}^{\infty} p(F_{i}|F)p(F|A,W_{0})dF
	\label{single}
\end{gather}
where $p(F_{i}|F)$ is a normal distribution centered in $F_{i}$ with uncertainty $\sigma_{i}$. Both values are measured by our line detection code for each object. We obtain $p(F|A,W_{0})$ assuming a normal continuum distribution with rest UV continuum flux $f_{\lambda,i}$ and uncertainty $\sigma_{\lambda, i}$:
\begin{gather}
\label{expansion}
p(F|A, W_{0})=\\
\nonumber
\int_{0}^{\infty} \frac{1}{|EW|}p(EW|A, W_{0})\frac{e^{(f_{\lambda,i}-F/EW)^{2}/2\sigma_{\lambda, i}^{2}}}{\sqrt{2\pi}\sigma_{\lambda, i}}dEW
\end{gather}
where $p(EW|A, W_{0})$ is the EW model given by equation (\ref{dist}), and both $f_{\lambda, i}$ and $\sigma_{\lambda, i}$ come from CANDELS photometry and our redshifts ($z_{Ly\alpha}$ or $z_{EAZY}$).

The limiting line flux $F^{*}_{i}$ for discerning detections from noise is given by our $SN$ threshold, i.e., $F^{*}_{i}=SN^{*}\sigma_{i}$. For galaxies with a detection that satisfies $F_{i}>F^{*}_{i}$, the single likelihood $p(F_{i}|A, W_{0})$ is determined by (\ref{single}). For galaxies with no detections above $F^{*}_{i}$, we adopt the value
\begin{gather}
\nonumber
	p(F_{i}<F^{*}_{i}|A,W_{0})= \\ \int_{0}^{\infty}\left(1-p(F_{i}>F^{*}_{i}|F)\right)p(F|A,W_{0})dF
	\label{nondet}
\end{gather}
with $p(F_{i}>F^{*}_{i}|F)$ our detection completeness at a line flux $F$. To obtain it, we characterize $p(SN_{i}>SN^{*}|SN)$ instead. We simulate $\sim 10^3$ lines on the 115 sky-spectra sampling fluxes of $10^{-17}-10^{-19}$ erg s$^{-1}$ cm$^{-2}$, FWHMs between $5$ \AA \ and $13$ \AA, and wavelengths of 4800-6700 \AA. 

Using the recovered expressions for detections and non-detections, the posterior distribution takes its final form:
\begin{gather}
\nonumber
p(A,W_{0}| \{F\})= \\  \frac{C}{AW_{0}} \displaystyle \prod_{D}p(F_{i}|A, W_{0}) \displaystyle \prod_{ND}p(F_{i}<F^{*}_{i}|A,W_{0})
\end{gather}

With the prior $p(A,W_{0}) \propto A^{-1} W_{0}^{-1}$. This is obtained by assuming $A$ and $W_{0}$ are independent and distribute uniformly in logarithmic scale. The constant $p(\{F\})$ represents the likelihood of the model. Hence, $C$ groups every constant so that $p(A,W_{0}| \{F\})$ integrates 1.

\section{Results}
We use a stellar mass selected sample of galaxies to derive the $3<z<4.6$ Ly$\alpha$ EW distribution. To study its dependence on $M_*$, we divide our sample in three bins covering the range $7.6<\log{M_*/M_{\odot}}<10.6$ (Figure 2). The observed distributions, along with our recovered models and constraints, are shown in Figure 3. The posterior distributions for $A$ and $W_{0}$ are presented in Figure 4. From these figures, we confirm that both parameters, the fraction $A$ of galaxies featuring Ly$\alpha$ emission and
the e-folding scale $W_{0}$ of the distribution, anti-correlate with $M_*$. To characterize this effect, we use linear parameterizations. We define the mass of each bin as its median mass and obtain
\begin{gather}
A(M_*) = -0.26_{-.11}^{+.13}\log{M_*/M_{\odot}}+ 3.01^{+1.0}_{-1.2} \\
W_{0} (M_*) = -15.6_{-3.5}^{+3.2} \log{M_*/M_{\odot}} + 166^{+34}_{-31}
\end{gather}
We also divide the whole sample in the two photometric redshift bins $3<z_{EAZY}<3.65$ and $3.65<z_{EAZY}<4.6$. We use the recalculated $M_*$ and $EW$, but select on $z_{EAZY}$ to avoid $\Delta z$ biases in the subsamples (Section 3.2). Then, we constrain $A(M_*)$ and $W_{0}(M_*)$ for both populations and find no significant differences (Figure 5).

\begin{figure}[h!]
	\centering
	\includegraphics[width=3.4in]{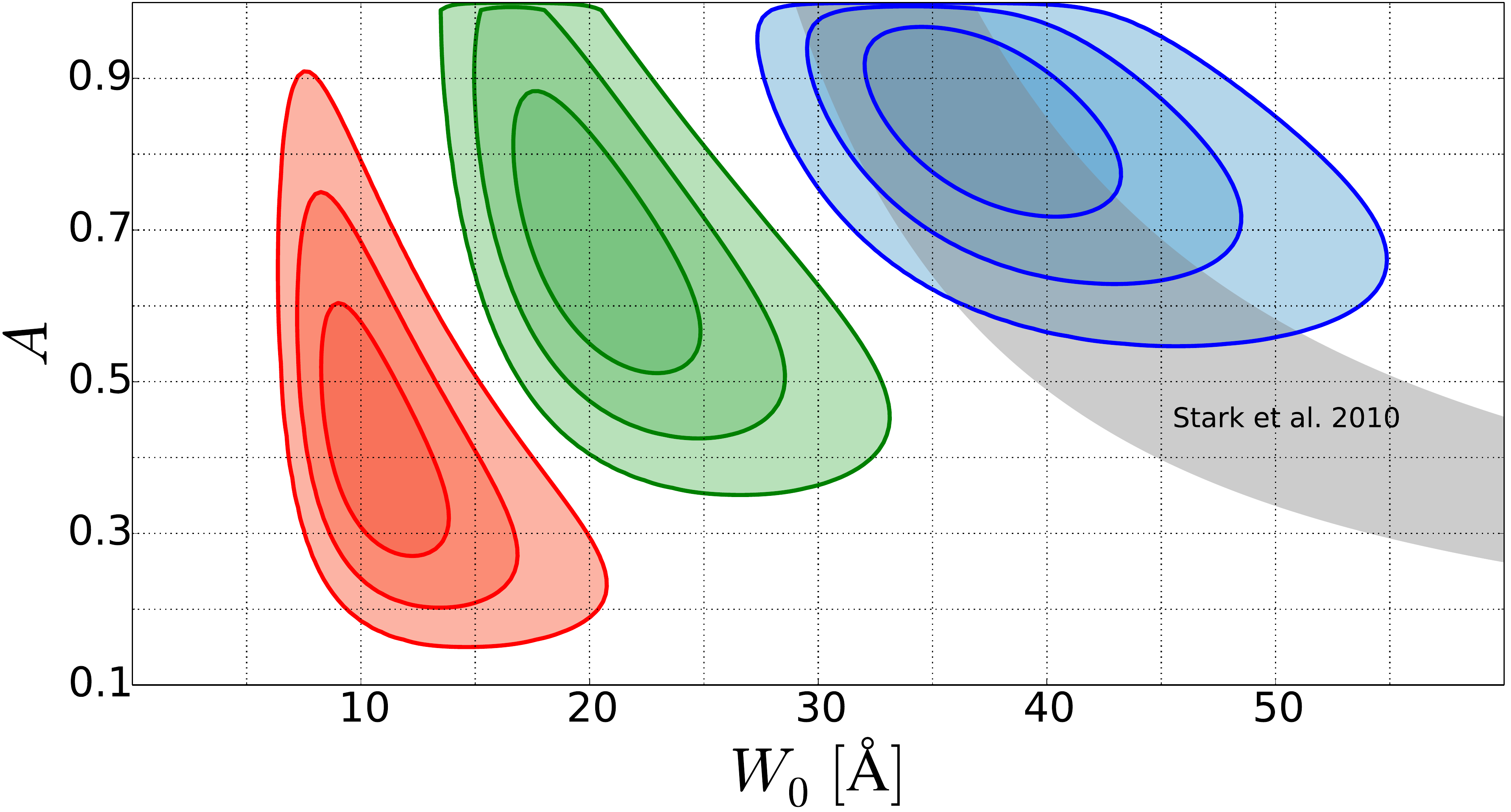}
	\caption{Low- (blue), medium- (green) and high-mass (red) posteriors for exponential parameters $A$ and $W_{0}$. The 3 contours for each subsample represent $1\sigma$, $2\sigma$, and $3\sigma$ confidence levels. The shaded region shows Stark et al.\ 2010 results on $z\sim 4$ LBGs.}
\end{figure}

\section{Summary and Discussion}
Using 3D-HST outputs, we design a $3<z<4.6$ sample of 629 galaxies in the range $7.6<\log{M_*/M_{\odot}}<10.6$. We conduct a spectroscopic survey of the 629 galaxies using M2FS, allowing us to measure Ly$\alpha$ fluxes. We measure the Ly$\alpha$ EW distribution for 3 different $M_*$ subsamples and model it using a Bayesian framework. We confirm an anti-correlation between $M_*$ and prominence of Ly$\alpha$ emission in galaxies, obtaining quantitative relations for the distribution parameters as a function of $M_*$. These relations are best reproduced by a low-mass population showing mostly emission and a high-mass counterpart where about half shows no emission/absorption.

Using $z\sim 4$ LBGs, Stark et al.\ 2010 find a $\sim10\%$ fraction of LAEs (EW $\geqslant 75$\AA). At $z\sim 4$, their sample $M_{UV}$ translates to $10^{8}-10^{10.5}M_{\odot}$ (Gonz\'alez et al.\ 2014). We simulate their selection in our data and find an $M_*$ distribution dominated by $10^{8}-10^{10}M_{\odot}$ objects. Thus, Figure 4 hints that Stark et al.\ 2010 results on the higher end of the EW distribution are dominated by $10^{8}-10^{9}M_{\odot}$ galaxies. Using a narrowband sample, Zheng et al.\ 2014 recover the $z\sim 4.5$ LAEs EW distribution. They find a best-fit $W_{0}=50\pm 11$ for EW$<400$\AA, but a much higher $W_{0}=167^{+44}_{-19}$ from simulations. Our results suggest that their composite EW distribution is a result of the broad $M_*$ range induced by narrowband surveys.

We measure a median $\Delta z= z_{Ly\alpha}-z_{EAZY}=0.24$. This offset apparently anti-correlates with $M_*$, i.e., correlates with EW (Figure 2). Therefore, we attribute this feature to Ly$\alpha$ line effects on EAZY fitting and will address it in future papers. Given $\Delta z$ and the median  $\sigma_{EAZY}=0.1$, we avoid any detailed analysis involving the broad redshift distribution of the sample. Nevertheless, the trends we recover are also observed when dividing the sample in two $z_{EAZY}$ bins (Figure 5).

While the methodology we present provides a Bayesian approach to deal with high-redshift Ly$\alpha$ emission statistics, the results allow for comparison between surveys with different mass sensitivity limits. These insights are essential for using Ly$\alpha$ statistics at different redshifts under the same scheme, allowing for proper interpretation of Ly$\alpha$ pre and post- reionization. In addition, the trends we recover also provide constraints for simulations, especially those devoted to statistically studying Ly$\alpha$ emission in the galaxy population (e.g.\ Zheng et al.\ 2010; Barnes et al.\ 2011).

\begin{figure}
	\centering
	\includegraphics[width=3.3in]{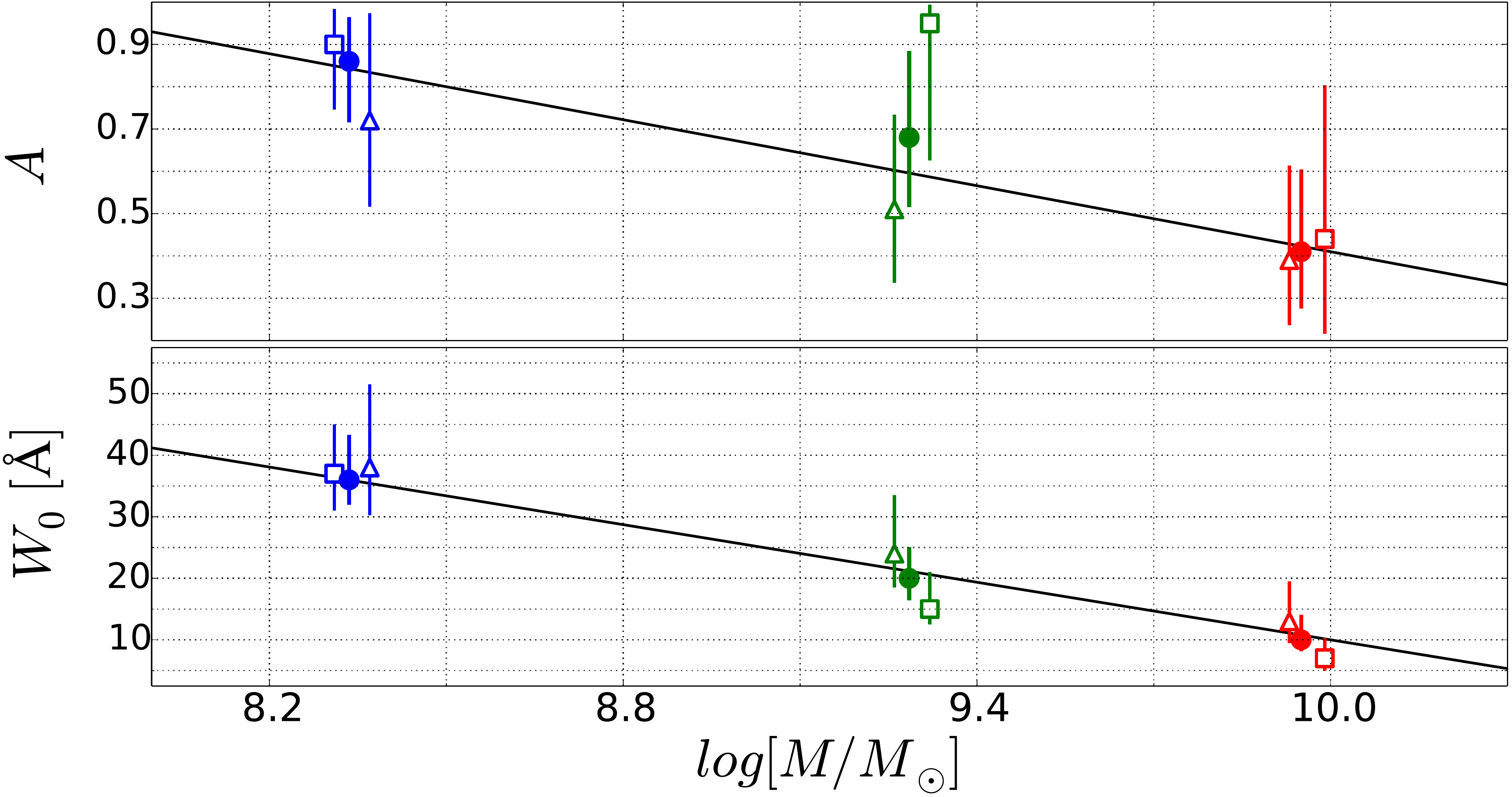}
	\caption{Exponential parameters $A$ (top) and $W_{0}$ (bottom) as a function of bin median $M_*$ for the complete sample (circles), $3<z_{EAZY}<3.65$ (triangles) and $3.65<z_{EAZY}<4.6$ (squares). The solid lines correspond to equations (9) and (10).}
\end{figure}


\acknowledgments
We thank Linhua Jiang and his team at Peking University for their contribution to improve our M2FS reduction pipeline. We also thank Andr\'es Escala for allowing us to make use of his computing cluster at Departamento de Astronom\'ia, Universidad de Chile. 

G.O. was supported by CONICYT, Beca Mag\'ister Nacional 2014, Folio 22140924. G.B. is supported by CONICYT/FONDECYT, Programa de Iniciaci\'on, Folio 11150220. E.O. was partially supported by NSF grant AST1313006. This work is based on observations taken by the 3D-HST Treasury Program (GO 12177 and 12328) with the NASA/ESA HST, which is operated by the Association of Universities for Research in Astronomy, Inc., under NASA contract NAS5-26555. This paper includes data gathered with the 6.5 m Magellan Telescopes located at Las Campanas Observatory, Chile. 

\clearpage


\clearpage







\clearpage


\begin{thebibliography}{}
\bibitem[Barnes et al.(2011)]{2011MNRAS.416.1723B} Barnes, L.~A., Haehnelt, 
M.~G., Tescari, E., \& Viel, M.\ 2011, \mnras, 416, 1723 		
\bibitem[Blanc et al.(2011)]{2011ApJ...736...31B} Blanc, G.~A., Adams, 
J.~J., Gebhardt, K., et al.\ 2011, \apj, 736, 31 
\bibitem[Brammer et al.(2008)]{2008ApJ...686.1503B} Brammer, G.~B., van 
Dokkum, P.~G., \& Coppi, P.\ 2008, \apj, 686, 1503 
\bibitem[Ciardullo et al.(2012)]{2012ApJ...744..110C} Ciardullo, R., 
Gronwall, C., Wolf, C., et al.\ 2012, \apj, 744, 110 
\bibitem[Charlot 
\& Fall(1993)]{1993ApJ...415..580C} Charlot, S., \& Fall, S.~M.\ 1993, \apj, 415, 580 
\bibitem[Gonz{\'a}lez et al.(2014)]{2014ApJ...781...34G} Gonz{\'a}lez, V., 
Bouwens, R., Illingworth, G., et al.\ 2014, \apj, 781, 34 
\bibitem[Grogin et al.(2011)]{2011ApJS..197...35G} Grogin, N.~A., Kocevski, 
D.~D., Faber, S.~M., et al.\ 2011, \apjs, 197, 35 
\bibitem[Hagen et al.(2014)]{2014ApJ...786...59H} Hagen, A., Ciardullo, R., 
Gronwall, C., et al.\ 2014, \apj, 786, 59 
\bibitem[Koekemoer et al.(2011)]{2011ApJS..197...36K} Koekemoer, A.~M., 
Faber, S.~M., Ferguson, H.~C., et al.\ 2011, \apjs, 197, 36
\bibitem[Kriek et al.(2011)]{2011ApJ...743..168K} Kriek, M., van Dokkum, 
P.~G., Whitaker, K.~E., et al.\ 2011, \apj, 743, 168 
\bibitem[Mateo et al.(2012)]{2012SPIE.8446E..4YM} Mateo, M., Bailey, J.~I., Crane, J., et al.\ 2012, \procspie, 8446, 84464Y
\bibitem[Ouchi et al.(2008)]{2008ApJS..176..301O} Ouchi, M., Shimasaku, K., 
Akiyama, M., et al.\ 2008, \apjs, 176, 301 
\bibitem[Schaerer(2003)]{2003A&A...397..527S} Schaerer, D.\ 2003, \aap, 397, 527
\bibitem[Shapley et al.(2003)]{2003ApJ...588...65S} Shapley, A.~E., 
Steidel, C.~C., Pettini, M., \& Adelberger, K.~L.\ 2003, \apj, 588, 65 
\bibitem[Skelton et al.(2014)]{2014ApJS..214...24S} Skelton, R.~E., 
Whitaker, K.~E., Momcheva, I.~G., et al.\ 2014, \apjs, 214, 24 
\bibitem[Stark et al.(2010)]{2010MNRAS.408.1628S} Stark, D.~P., Ellis, 
R.~S., Chiu, K., Ouchi, M., \& Bunker, A.\ 2010, \mnras, 408, 1628 
Bogosavljevi{\'c}, M., Shapley, A.~E., et al.\ 2011, \apj, 736, 160 
\bibitem[Verhamme et 
al.(2006)]{2006A&A...460..397V} Verhamme, A., Schaerer, D., \& Maselli, A.\ 2006, \aap, 460, 397 
\bibitem[Zheng et al.(2014)]{2014MNRAS.439.1101Z} Zheng, Z.-Y., Wang, 
J.-X., Malhotra, S., et al.\ 2014, \mnras, 439, 1101 
\bibitem[Zheng et al.(2010)]{2010ApJ...716..574Z} Zheng, Z., Cen, R., Trac, 
H., \& Miralda-Escud{\'e}, J.\ 2010, \apj, 716, 574 
\end{thebibliography}
\end{document}